# Cooperation Dynamics in Multi-Agent Systems: Exploring Game-Theoretic Scenarios with Mean-Field Equilibria[*]

Sathi Vaigarai[†], Sabahat Shaik[†], and Jaswanth Nidamanuri[†]

*Abstract*— Cooperation is fundamental in Multi-Agent Systems (MAS) and Multi-Agent Reinforcement Learning (MARL), often requiring agents to balance individual gains with collective rewards. In this regard, this paper aims to investigate strategies to invoke cooperation in game-theoretic scenarios, namely the iterated prisoner's dilemma, where agents must optimize both individual and group outcomes. Existing cooperative strategies are analyzed for their effectiveness in promoting group-oriented behavior in repeated games. Modifications are proposed where encouraging group rewards will also result in a higher individual gain, addressing real-world dilemmas seen in distributed systems. The study extends to scenarios with exponentially growing agent populations ($N\rightarrow+\infty$), where traditional computation and equilibrium determination are challenging. Leveraging mean-field game theory, equilibrium solutions and reward structures are established for infinitely large agent sets in a model-based scenario. Finally, practical insights are offered through simulations using the Multi Agent – Posthumous Credit Assignment trainer, and the paper explores adapting simulation algorithms to create scenarios favoring cooperation for group rewards. These practical implementations bridge theoretical concepts with real-world applications.

*Keywords— Game Theory, Iterated Prisoner's Dilemma, Mean-field Game, Multi-Agent Cooperation, Multi-Agent Reinforcement Learning, Multi-Agent Systems.*

## I. Introduction

Cooperation is a cornerstone in the realm of Multi-Agent Systems (MAS) and Multi-Agent Reinforcement Learning (MARL), playing a pivotal role in scenarios where agents must strike a delicate balance between individual gains and collective rewards. While off-the-shelf MARL algorithms have historically navigated complex dynamic interactions in multi-agent environments, the rising prominence of model-based MARL is evident due to its increased utility and potential for enhanced real-world tasks [1]. With MARL's ubiquity in real-world use cases like autonomous vehicles [2], healthcare [3], and broader game theory applications [4], delving into these algorithms and exploring novel strategies for refining their efficiency becomes imperative. Curse of dimensionality is a challenge when dealing with scenarios where $N >> 2$, but when $N \rightarrow +\infty$, the learning becomes tractable via mean-field approximation.

In parallel, the simulation and validation of such environments, underpinned by theoretical concepts, have spurred the development of various algorithms that cover diverse facets of MARL. One such effort is from Unity Technologies [5], who introduced an intriguing addition to their ML-Agents toolkit—the MA-POCA (Multi-Agent Posthumous Credit Assignment) trainer. This effort aims to address challenges in nurturing agents' comprehension of their group contributions, even in the event of their demise and subsequent low rewards [6]. By encouraging agents to prioritize the group's welfare over the individual gain, MA-POCA presents a compelling approach. However, the adaptability of decision-making scenarios to harmonize individual objectives with group rewards warrants exploration, optimizing alignment between the two.

This work aims to make the following contributions:

(i). Investigate strategies for instigating cooperation within an iterated prisoner's dilemma while optimizing agent interests.

(ii). Extend this principle to an NIPD (N-player Iterated Prisoner's Dilemma), formulating optimal reward structures and equilibrium strategies through the lens of mean-field game theory.

(iii). Offer practical insights into the simulation of such environments, analyze existing MARL algorithms, and explore the scope for adapting them into the dynamic scenarios as envisioned above.

The subsequent sections are organized as follows: Section II surveys the prior works in the literature which address the challenges and goals similar to that of this study. Section III provides a comprehensive exposition of the proposed model-based approaches and highlights the underlying mathematical foundations behind these approaches and their anticipated functionality. Section IV delves into the results, showcasing how these theoretically derived strategies translate into practical applications, and examines the explored algorithms underpinning them. Section V concludes this work, offering insights into the challenges not addressed by this study and charting promising directions for future research.

## II. Related work

In recent years, the intersection of game theoretic models and reinforcement learning has garnered attention for optimization purposes. A notable contribution by authors [7] uses the Stackelberg game structure, offering enhancements to the traditional actor-critic-based reinforcement learning. Meanwhile, another study [8] delved into equilibrium scenarios within the iterated prisoner's dilemma using various game-theoretic approaches, resulting in optimal strategies. However, this work primarily focuses on memory-one strategies and confines its analysis to scenarios where the inequality $2R > T + S$ holds true. In contrast, [9] explored an N-player iterated prisoner's dilemma, proposing dynamic interconnected topologies to foster cooperation within the N-player system. Furthermore, [10] introduced a novel approach employing signed networks to establish negative links between players in the game, effectively promoting cooperation. It's worth noting that this methodology is tailored to the two-player version of the game, rendering it less suitable for the more complex N-player iterative version where invoking cooperative behavior becomes more challenging.

[†] School of Technology, Woxsen University, Hyderabad - 502345, India. {sathi.vaigarai_2024, sabahat.shaik_2024, jaswanth.nidamanuri}@woxsen.edu.in

Notably, existing literature tends to explore the treatment of repeated game scenarios as mean-field games to a lesser extent. The authors in [11] highlight several topics of interest in MARL from a game theoretical perspective, including learning in zero-sum games, general-sum games, and the inclusion of mean-field when the number of agents is large. Studying equilibrium scenarios within such dynamic environments holds a significant promise for real-time applications. The following section outlines our proposed approach, initially deriving equilibrium strategies in scenarios not requiring a mean-field approach. Subsequently, this approach is extended to a mean-field scenario, thereby scaling the solution to identify optimal strategies crucial for achieving equilibrium. This approach contributes to the burgeoning field of game-theoretic reinforcement learning, while paving way for practical applications in dynamic, multi-agent environments.

## II. Proposed Methodology

This section highlights the proposed strategic approaches towards the different scenarios considered—Iterated Prisoner's Dilemma, and the N-Player Iterated Prisoner's Dilemma using a model scenario. A detailed overview of the mathematical foundations behind these strategies, and how they are expected to influence agents' behavior in comparison to current strategies such as Win-Stay Lose-Shift and Grim strategy are explored, and further discusses the approaches which can be taken in the formulation of reinforcement learning algorithmic strategies which would be crucial in the simulation of such environments.

### A. Iterated Prisoner's Dilemma

The normal form of the prisoner's dilemma and the iterated version is provided in Table 1.

TABLE I. PRISONER'S DILEMMA PAYOFF MATRIX

| X \ Y | Cooperate | Defect |
|---|---|---|
| Cooperate | R, R | S, T |
| Defect | T, S | P, P |

R is the reward to each player for cooperation, T is the temptation payoff for the defector when the other player cooperates, S is the sucker's payoff for cooperation when the other player defects and P is the punishment when both players defect. To ensure the nature of the game, the inequalities $T > R > P > S$ and $2R > T + S$ should be met. This results in the agents being forced to cooperate, as mutual cooperation is better than defection. But this means that each agent must sacrifice getting the maximum reward in each iteration to ensure the betterment of the group. Here, a new strategy is proposed, by crafting a scenario where the cooperation group reward is lesser than the betrayal scenario where one cooperates and other defects, i.e., $2R < T + S$, the agents can be encouraged to take turns obtaining the maximum reward for $i$ iterations, without sacrificing the group reward. In this scenario, if the current choice tuple is (C, D), yielding a reward of (S, T), and the agents agree to alternatively defect, then the strategy for the next iteration can be (D, C), yielding a reward of (T, S). Each agent thus takes turns obtaining the maximum reward while also ensuring that the group reward is maximized.

An agent who has chosen to defect in the current iteration will not deviate from the strategy and defect in the next iteration, as it will not only provide it the lowest possible reward, but in the high chance that the other agent sticks to the agreed strategy, the group reward will also be minimum. But an agent who has cooperated in the current iteration might cooperate in the next, where the other agent's and the group's reward will be minimized. If the other agent stuck to the strategy and was denied maximum reward, then it may retaliate by defecting forever to ensure that its reward will always be greater than or equal to the first agent. This would result in both obtaining the punishment reward, until such a time when they go back to their original strategy.

For the agents to not deviate from the strategy, the reward obtained from not deviating from the strategy (alternating between maximum and minimum reward) must be higher than the reward obtained for deviating (maximum reward in current iteration, followed by punishment reward in all others). Considering the discount factor $\delta$ for future iterations:

- Reward if agent sticks to strategy:
$$T + S\delta + T\delta^2 + S\delta^3 + T\delta^4 + \ldots = \frac{T+S\delta}{1-\delta^2} \quad (1)$$

- Reward if agent deviates from strategy:
$$T + P\delta + P\delta^2 + P\delta^3 + P\delta^4 + \ldots = T + \frac{P\delta}{1-\delta} \quad (2)$$

For agent to not deviate from strategy, (1) > (2):
$$\frac{T+S\delta}{1-\delta^2} > T + \frac{P\delta}{1-\delta}$$

Which results in the condition:
$$\delta > \frac{P-S}{T-R} \quad (3)$$

If the discount factor is above this level, cooperation by continually alternating mutual sacrifice of reward can be achieved.

In the Grim Trigger strategy, cooperation is maintained until one player defects, at which point both players defect indefinitely. In contrast, the proposed strategy allows for occasional cooperation by taking turns maximizing rewards, potentially leading to more dynamic and variable outcomes. On the other hand, the Win-Stay Lose-Shift strategy involves staying with the chosen action if it leads to a win and shifting to the opposite if it results in a loss. The proposed methodology is essentially a Win-Shift Lose-Shift strategy, as players shift strategy regardless of the outcome, albeit ensuring that neither picks the same choice at any given iteration. This highlights the importance of a dynamic approach to cooperation, where agents adapt their strategies based on recent interactions to ensure fairness and maximize their own gains over time.

## B. Mean-Field Equilibria

**Model Scenario**: Imagine an intersection with a large population of $N \to +\infty$ agents, where each agent is a vehicle waiting to cross the intersection. Each agent has two choices: wait or move. However, the dynamics are influenced by the number of agents who chose to move and a threshold $i$ ($0 < i < N$), that limits the maximum number of agents allowed to pass ahead. If the number of agents who chose to move $j \leq i$, then all $j$ agents will make it through, and $N-j$ agents will wait. In case $j > i$, then all the $N$ agents will not move. The scenario then repeats with the agents being provided the same choices, with knowledge of the previous iteration's outcome.

*Game Framework*: The dynamic plays out as a repeated game, a multi-agent extension of the classic N-player iterated prisoner's dilemma where cooperation and defection are not individual choices but a group outcome, which is infused with the added complexity of mean-field scenarios. Within this backdrop, the primary objective is to craft a reward system and equilibrium scenarios that ensure, in each iteration, a minimum of $j$ vehicles traverse the intersection, and that $j \to i$ is achieved. The parameters underpinning this formulation aim to incentivize altruistic behavior, secure individual rewards for each agent's contribution, foster cooperation while motivating agents to pursue individual rewards, and ultimately maximize collective rewards. Approaching this as a discrete-time mean-field game, several mathematical facets come into play.

*States and Action*: We define the state variable $S$ as $j$, representing the number of agents who opted to move in the preceding iteration. It forms a finite set encompassing all feasible values of $j$, ranging from 0 to $N$. Agents, equipped with their knowledge of the current state and their past decisions, navigate the current iteration with a binary action from the action space $a$: 0 for 'wait' and 1 for 'move.'

*Reward Structure*: The reward structure delineates distinct scenarios:

- If $j \leq i$ (Cooperative Case):
  1. Agents who move receive a good reward for moving.
  2. Agents who wait receive a good reward for their cooperative choice.
- If $j > i$ (Defection Case):
  1. Agents who move receive the least for moving.
  2. Agents who wait receive a low reward for not contributing to overcrowding.

In essence, agents reap the highest rewards for actions that minimize the number of agents making the same choice. However, they must also strive to maximize the collective reward, achievable through equilibrium scenarios. It's worth noting that deviating from cooperation after equilibrium is reached invariably proves detrimental to the agent.

Consequently, the group reward $R$ is formulated as:

$$R = \sum_{k=0}^{N} P(j) \cdot \frac{1}{1 + e^{(1-2a_k)(i-j)}} + B$$

Here, $a_k$ signifies the action taken by agent $k$, $P(j,t)$ denotes the fraction of agents in state $j$ at time $t$, and $B$ serves as a constant offset.

Utility Function: To incentivize cooperative choices, the utility function is formulated to incorporate the disparity $j-i$, signifying the proximity or divergence from equilibrium. The utility function takes the form:

$$U[j] = -a|j - i| + b$$

In this utility function:

- $a$ represents a parameter influencing the agent's inclination for consistency when $j$ approaches the threshold $i$, signifying the agent's preference for decisions harmonizing with the threshold. Higher $a$ values denote a stronger inclination for consistency.

- $b$ serves as a constant factor representing the agent's preferences, encapsulating additional factors beyond the $j-i$ difference. $b$ may embody a baseline preference for one choice, independent of the threshold.

This utility function captures the interplay between individual preferences and the collective pursuit of equilibrium.

*Mean-Field Distribution Evolution*: Agents make decisions considering the average behavior of the entire population. Here, the state variable $j$ represents the average behavior of all agents in the population. Agents consider the collective impact of their decisions on the overall state, and this influence affects their individual choices. We find the mean-field distribution $P(j,t)$, which represents the fraction of agents occupying a state $j$ at time $t$. In this scenario, $j$ is the key state variable, and the distribution evolves over time as agents make decisions. At each time step $t$, $P(j,t)$ should capture the probability that $j$ agents have chosen to move in the current iteration, based on the previous iterations' outcomes and the agents' strategies.

To capture the evolution of the mean-field distribution, we define how $P(j,t+1)$ depends on $P(j',t)$ at time $t$ and the choices of actions made by agents. In this case, the evolution is described as follows:

$$P(j,t+1) = \sum_{a=0}^{1} \sum_{j'=0}^{N} T(j',j,a,t) \cdot P(j',t) \cdot \pi(a,j',t)$$

Where $T(j',j,a,t)$ is the transition probability from state $j$ at time $t$ to state $j$ at time $t+1$ given action $a$; $\pi(a,j',t)$ describes the probability that an agent chooses action $a$ at time $t$ given the current state $j'$. The double summation accounts for all possible actions and states of agents that could lead to the state $j$ at time $t+1$. The probability $\pi(a,j',t)$ is computed using a SoftMax function based on the agents' value functions:

$$\pi(a,j',t) = \frac{\exp(V(a,j',t))}{\sum_{a'=0}^{1} \exp(V(a',j',t))}$$

By iteratively applying this update rule for *P(j,t)* over time, we can track how the distribution of states evolves as agents make decisions in each iteration. The equilibrium, in this case, would be reached when the distribution stabilizes, and agents' actions are unlikely to change further as they found strategies which maximize their expected utility given the distribution and rewards associated with their actions.

*Bellman Equation*: The Bellman optimality equation represents the optimal value of being in a particular state in a Markov Decision Process (MDP), and the general form of the equation for a single agent can written as:

$$V^*(s) = max_a \sum_{s'} P(s'|s,a)[R(s,a,s') + \delta V^*(s')]$$

Where *V\*(s)* is the value function representing the expected cumulative reward of an agent being in state *s* and following an optimal strategy; *a* is the action taken in state s to maximize the expected cumulative reward; *P(s'|s,a)* is the transition probability that the MDP will transition from a state *s* to state *s'* when taking action *a*; *R(s, a, s')* is the immediate reward received when transitioning states; *δ* is the discount factor, representing the preference for immediate rewards over future rewards. This seeks to find the action *a* which maximizes the expected cumulative reward, thereby optimizing the actions of the agent.

In a discrete-time mean-field game, this is used to find the optimal strategy for a rational agent in a population where each agent's behavior influences and is influenced by the average behavior of the population. In a population of *N* agents indexed by *i*, let $x_i(t)$ denote the state of agent *i* at time *t* and *m(t)* denote the average behavior of the population, the Bellman equation for a representative agent in this framework can be written as:

$$V(x_i(t), m(t)) = max_{u_i(t)} \{J_i(x_i(t), u_i(t), m(t)) + \delta \mathbb{E}[V(x_i(t+1), m(t+1))]\}$$

Where $J_i(x_i(t), u_i(t), m(t))$ is the immediate utility based on its own state, strategy, and the current mean field. $E[V(x_i(t+1), m(t+1))]$ represents the expected future value for agent *i* at time *t+1*, given the transition dynamics and the anticipated mean field at the next time step *m(t+1)*, which essentially captures the long-term objectives of the agent. Consequently, for the described model in this section, the Bellman optimality equation can be written as:

$$V(j,t) = max_{a \in \{0,1\}} \{U(a,j) + \delta \mathbb{E}[V(j', t+1)]\}$$

Where *U(a,j)* is the utility function for an agent who chooses action *a* (0 for wait, 1 for move), given the current state *j*. And *E[V(j', t+1)]* is the expected value of the value function in the next time step *t+1*, given the transition dynamics of *j* and agent strategies.

Overall, this scenario presents a sophisticated framework for studying multi-agent dynamics and equilibrium in the context of a simple model, where agents must balance their individual preferences with the goal of achieving a collective equilibrium. The mathematical formulations applied to the model to analyze the dynamics shed light on how agents can optimize their strategies in this complex environment, and how cooperation can be induced with the help of simple reward and equilibrium strategies.

*C. Posthumous Credit Assignment*

Although the Posthumous Credit Assignment trainer (Cohen et al. 2021) was developed specifically for purposes of letting agents know when the group has benefited from their actions in the event of the agents being terminated, this work considers it from the perspective of providing delayed rewards to the agent, when it is still active, but not performing a task for the benefit of the group. To address it from the scenario described in the N-player iterated prisoner's dilemma, an agent who is waiting to let the other agents move will be provided a delayed reward based on how their actions contributed towards the performance of the whole group. For reasons of consistency, this will be referred to as posthumous credit assignment.

One observation regarding this methodology is the stagnation of roles of agents over a period, and reduced dynamism in decisions. While this can be beneficial when all the individual agents are just there to serve the purposes and goals of the larger collective, but in real-world scenarios, each agent will be in a position to prioritize their individual goals as much as they do the group ones. In context of the mean-field scenario provided earlier, this would mean that the application of the current credit assignment policy would result in eventually reaching equilibrium where $j \rightarrow i$, but in an environment of *N* agents, the over many iterations, the same *j* agents would choose to move while the same *N–j* agents choose to wait. Given that the agents receive the maximum reward when they move ($j \leq i$), real-world scenarios would involve each agent wanting to obtain the maximum reward.

A workaround to this can be obtained by dynamically interchanging the roles of the agents once stability is achieved, i.e., an agent who chose to wait can move in the next iteration while an agent who chose to move can wait in the next iteration. This ensures that the group reward is maximized as it was earlier, but in addition, all agents get to experience maximum rewards.

If the number of agents is two, as in a traditional iterated prisoner's dilemma, this can be implemented by implementing a simple memory-one strategy for the agents in the environment, where they know the actions taken in the prior iteration and can thus modify their strategies to let another agent get a shot at the maximum reward. Cooperation is utmost, as it involves an agent who has obtained maximum reward sacrificing it for a lower reward for the benefit of another agent, a behavior which can also be rewarded after the completion of successful iterations to encourage agents to develop this practice. Nevertheless, as the number of agents increases, the number of prior iterations to be considered before the choices for the current iteration can be made also increases significantly. Although the memory-one strategy will be sufficient for the experiments discussed in the following section, Section V provides possible ways to handle dynamic role switching where *N >> 2*.

III. RESULTS

In this section, we evaluate the performance of MA-POCA with the modifications proposed earlier for dynamic role switching empirically on a custom multi-agent environment like the one described in Section III (A), and an existing environment which was originally used to study the performance of MA-POCA trainer. All environments were developed using Unity's ML-Agents toolkit. Code is available at https://github.com/dawnorak/marl-kart-simulator.

### A. Iterated Prisoner's Dilemma

A simple environment is constructed where the goal of the agent is to go around a track on a kart. The implementation of the iterated prisoner's dilemma comes when the agents approach an intersection, where only one is permitted to move ahead. The rewards are then provided based on the standard prisoner's dilemma payoff matrix, but where $2R < T + S$, as proposed in Section III, to encourage the agents to alternatively shift strategies and take turns going around the track. This would further need the implementation of a memory strategy where the agents know the choice they made prior and shift accordingly. Due to the simplicity of the simulation, the implementation of the discount factor to ensure agents do not deviate from strategy was not needed. The implementation of this strategy in the custom simulation is shown in Fig 1.

mainly, to remember whether they sacrificed themselves to obtain the key or not. This needed to be done only for the past two iterations, as in an $N$ agent system, the history of $N-1$ iterations can provide an idea of the choice to be made in the current iteration. In case an agent has sacrificed itself in any one of the past two iterations, then it will not sacrifice itself in the current iteration. If an agent has not sacrificed itself in the past two iterations and escaped, then it will look to sacrifice itself in the current iteration.

On modifying the agents' behavior to include their actions in the previous iterations—thereby encouraging the agents to take turns to sacrifice themselves—a similar level of success in the completion of objectives was achieved, albeit with the newer dynamic role shifting. Such a memory strategy could also be extended to other environments where dynamic role switching is needed. The implementation of this in the Dungeon Escape environment is shown in Fig. 2.

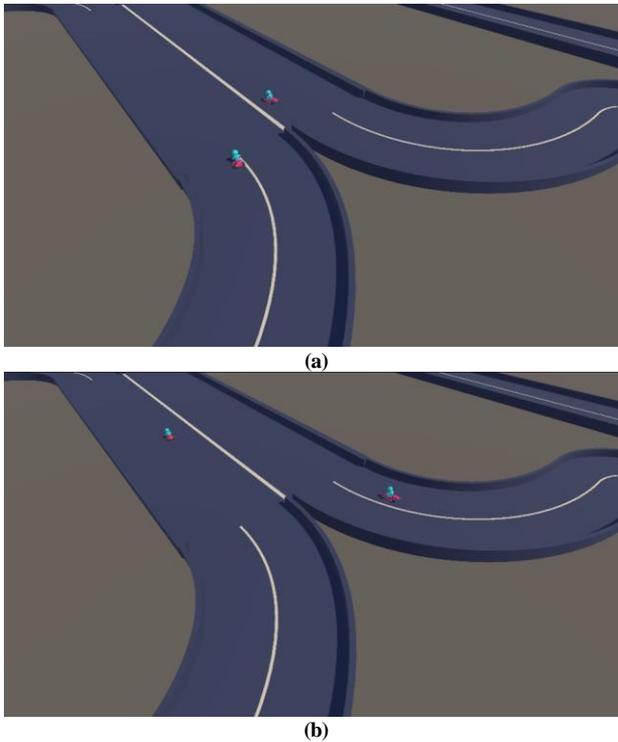

Figure 1: **Karts at an Intersection** – **(a)** Agent 1 waits while Agent 2 moves in the first iteration, **(b)** Agent 2 waits in the next iteration while Agent 1 moves.

### B. Dungeon Escape

The aim of this environment is for the blue agents to kill the green dragon by any one agent sacrificing itself, obtain the revealed key, and escape the dungeon. While training the environment as it was provided in the toolkit, it was observed that over iterations, one agent tends to become the sacrifice often, while the other agents escape. While this is efficient given that this environment is primarily to prioritize team goals over individual ones, this can be further explored to check the validity of using memory strategies to dynamically facilitate in the shifting of agent roles over iterations.

The three agents were given an additional task of remembering what their actions were in the previous iterations,

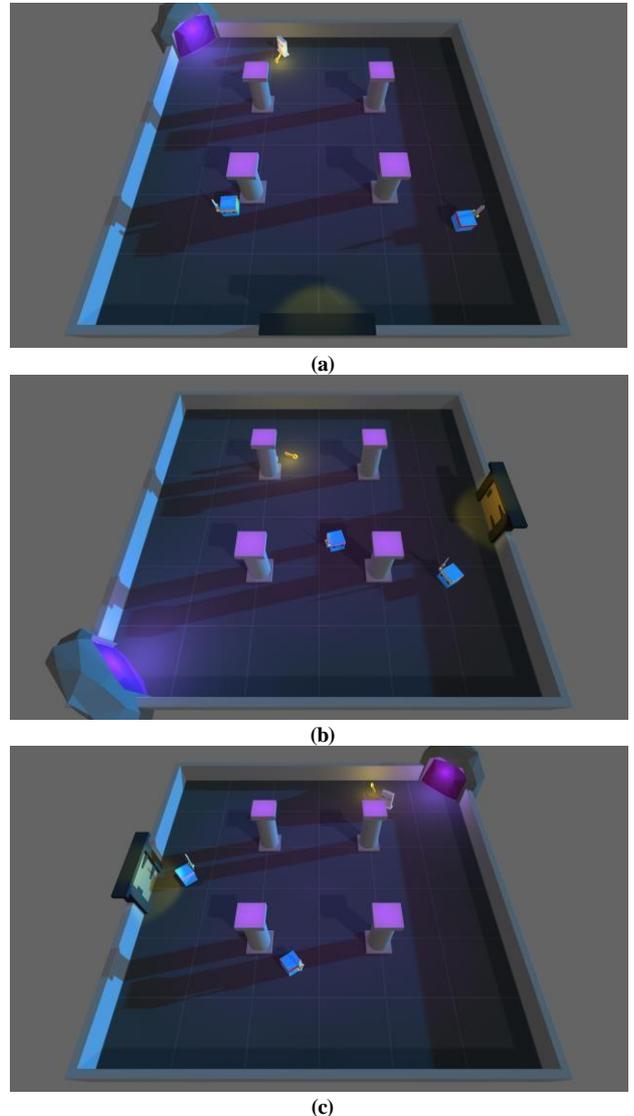

Figure 2: **Dungeon Escape** – **(a)** Purple headband agent sacrifices itself in iteration 1, **(b)** Red headband agent sacrifices itself in iteration 2, **(c)** Yellow headband agent sacrifices itself in iteration 3.

## IV. Conclusion

In summary, this work has introduced a few novel strategies for addressing game-theoretic scenarios and has developed corresponding reward policies and equilibrium concepts, particularly when dealing with scenarios where $N \to +\infty$ through a mean-field game perspective. These theoretical approaches have provided valuable insights into the mathematical and algorithmic dimensions of multi-agent reinforcement learning.

While our experimentation with MA-POCA in this work has yielded promising results in scenarios involving a small number of agents, the challenges become significantly intricate when dealing with scenarios characterized by $N$ being very large. In such cases, the extensive number of iterations required for dynamic role shifting presents a formidable challenge.

One potential avenue for tackling this challenge involves introducing stochastic decision-making and employing a probabilistic approach to role switching. Agents could maintain awareness of their cumulative rewards over numerous iterations, allowing them to gauge how long they have adhered to a specific strategy. For instance, if an agent has been sticking to the same strategy for $^{N-1}C_i$ (all possible combinations of $i$ agents in the agent space which does not consider the agent itself) iterations, that would mean that other agents have had ample opportunities to choose the other strategy. Consequently, the agent may decide to switch strategies based on a calculated probability. This probability can be expressed in any straightforward function, such as a sigmoid function, to represent the likelihood that an agent will transition to a different strategy at any given iteration. This stochastic approach offers a promising direction for addressing scenarios with a vast number of agents, promoting adaptability and strategic diversity in multi-agent systems.